\documentstyle[emulateapj5,epsf]{aastex}
\def \vt{\vartheta}

\shortauthors{Hoekstra et al.}
\shorttitle{Study of galaxy biasing}
\begin{document}

\title{Weak lensing study of galaxy biasing}

\author{Henk~Hoekstra$^{1,2,3}$, Ludovic~van Waerbeke$^{4,1}$,
Michael D.~Gladders$^{2,3,5}$, Yannick Mellier$^{4,6,3}$, and
H.K.C.~Yee$^{2,3}$}

\begin{abstract}

We combine weak lensing measurements from the Red-Sequence Cluster
Survey (RCS) and the VIRMOS-DESCART survey, and present the first
direct measurements of the bias parameter $b$ and the galaxy-mass
cross-correlation coefficient $r$ on scales ranging from 0.2 to 9.3
$h_{50}^{-1}$ Mpc (which correspond to aperture radii of $1.5'-45'$)
at a lens redshift $z\simeq 0.35$. We find strong evidence that both
$b$ and $r$ change with scale for our sample of lens galaxies
($19.5<R_C<21$), which have luminosities around $L_*$. For the
currently favored cosmology $(\Omega_m=0.3,~\Omega_\Lambda=0.7)$, we
find $b=0.71^{+0.06}_{-0.04}$ (68\% confidence) on a scale of
$1-2h_{50}^{-1}$ Mpc, increasing to $\sim 1$ on larger scales.  The
value of $r$ has only minimal dependence on the assumed cosmology.
The variation of $r$ with scale is very similar to that of $b$, and
reaches a minimum value of $r\sim 0.57^{+0.08}_{-0.07}$ (at $1
h_{50}^{-1}$ Mpc; 68\% confidence). This suggests significant
stochastic biasing and/or non-linear biasing. On scales larger than
$\sim 4 h_{50}^{-1}$ Mpc the value of $r$ is consistent with a value
of $r=1$.  In addition we use RCS data alone to measure the ratio
$b/r$ on scale ranging from 0.15 to 12.5 $h_{50}^{-1}$ Mpc ($1'-60'$)
and find that the ratio varies somewhat with scale. We obtain an
average value of $b/r=1.090\pm0.035$, in good agreement with previous
estimates.  A (future) careful comparison of our results with models
of galaxy formation can provide unique constraints, as $r$ is linked
intimately to the details of galaxy formation.

\end{abstract}

\keywords{cosmology: observations $-$ dark matter $-$ gravitational lensing}

\section{Introduction}

The growth of structures in the universe via gravitational instability
is an important ingredient in our understanding of galaxy formation.
However, the connection to observations is not straightforward, as we
need to understand the relation between the dark matter distribution
and the galaxies themselves. Galaxy formation is a complex process,
and it is not guaranteed a priori that this relation, referred to as
galaxy biasing, is a simple one. The bias might be non-linear, scale
dependent or stochastic. In the simplest case, linear deterministic
biasing, the relation between the dark matter and the galaxies can be
characterized by a single number $b$ (e.g., Kaiser 1987).

Most observational constraints of biasing come from dynamical studies
(see Strauss \& Willick 1995) which probe relatively large scales ($10
h_{50}^{-1}$~Mpc or more). Recent estimates on these scales suggest
values of $b\sim 1$ for $L_*$ galaxies (e.g., Peacock et al 2001;
Verde et al. 2001). On smaller scales some constraints come from
measurements of the galaxy two-point correlation function, which is
compared to the (dark) matter correlation function computed from
numerical simulations.  

\vbox{
\vspace{0.5cm}
\footnotesize
\noindent 
$^1$~CITA, University of Toronto,
Toronto, Ontario M5S 3H8, Canada\\
$^2$~Department of Astronomy and Astrophysics, University of Toronto,
Toronto, Ontario M5S 3H8, Canada; hoekstra,gladders,hyee@astro.utoronto.ca\\
$^{3}$~Visiting Astronomer, Canada-France-Hawaii Telescope, which
is operated by the National Research Council of Canada, Le Centre 
National de Recherche Scientifique, and the University of Hawaii\\
$^{4}$~Institut d'Astrophysique de Paris, 98 bis, Boulevard Arago, 75014,
Paris, France\\
$^{5}$~Present address: Observatories of the Carnegie Institution of 
Washington, 813 Santa Barbara Street, Pasadena, California 91101\\
$^{6}$~Observatoire de Paris, LERMA, 61, Avenue de l'Observatoire, 75014,
Paris, France
}

These studies indicate that the bias parameter $b\simeq 0.7$ on scales
less than $\sim 2 h_{50}^{-1}$ Mpc
$(\Omega_m=0.3,~\Omega_\Lambda=0.7)$. On larger scales $b$ increases
to a value close to unity (Jenkins et al.  1998).

Although this procedure provides useful information about the bias
parameter $b$ as a function of scale, it does rely on the assumptions
made in the numerical simulations. In addition, it cannot be used to
examine how tight the correlation between the matter and light
distribution is. To do so, we need to measure the galaxy-mass
cross-correlation coefficient $r$, which is a measure of the amount of
stochastic and non-linear biasing (e.g., Pen 1998; Dekel \& Lahav
1999; Somerville et al. 2001).

The coherent distortions in the shapes of distant galaxies caused by
weak gravitational lensing provide a unique way to study the dark
matter distribution in the universe. Weak lensing probes the dark
matter distribution directly, regardless of the light distribution. In
addition, it provides measurements on scales from the quasi-linear to
the non-linear regime, where comparisons between observations and
predictions are still limited. Much progress has been made in 
recent years, and the latest results give accurate joint constraints
on the matter density $\Omega_m$, and the normalisation of the
power spectrum $\sigma_8$ (e.g., Bacon et al. 2002; Hoekstra et al. 2002b;
Refregier et al. 2002; van Waerbeke et al. 2002).

Although redshift surveys can be used to determine the relative values
of $b$ and $r$ for different galaxy types (Tegmark \& Bromley 1999;
Blanton 2000), weak gravitational lensing provides the only direct way
to measure the galaxy-mass cross-correlation function (e.g., Fischer
et al. 2000; Wilson, Kaiser, \& Luppino 2001; McKay et al. 2001;
Hoekstra, Yee \& Gladders 2001b).  Fischer et al. (2000) used the SDSS
commissioning data to measure the galaxy-mass correlation function and
their results suggested an average value of $b/r\sim 1$ on
submegaparsec scales. This approach has been explored by Guzik \&
Seljak (2001) who used semi-analytic models of galaxy formation
combined with N-body simulations.  Their results suggest that the
cross-correlation coefficient is close to unity. The galaxies used in
their analysis, however, are massive (and consequently luminous)
because of the limited mass resolution of the simulations.

In this Paper we use the method proposed by Schneider (1998) and Van
Waerbeke (1998), which allows us to measure the biasing parameters
directly as a function of scale. Hoekstra, Yee, \& Gladders (2001a)
applied this technique to 16 deg$^2$ of $R_C$-band imaging data from
the Red-Sequence Cluster Survey (RCS; e.g., Yee \& Gladders 2001), and
measured the ratio $b/r$ as a function of scale. They found an average
value of $b/r=1.05^{+0.12}_{-0.10}$ for a $\Lambda$CDM cosmology on
scales ranging from $150h_{50}^{-1}$~kpc to $3h_{50}^{-1}$~Mpc.

However, an accurate measurement of the mass auto-correlation function
is required to constrain $b$ and $r$ separately. Furthermore the
analysis is facilitated if the measurements of both the galaxy
auto-correlation function and the mass auto-correlation function probe
the power spectrum at the same redshift. To this end, the measurements
presented by van Waerbeke et al. (2002), based on deep imaging data
from the VIRMOS-DESCART survey, are ideal. In this Paper we combine
the results of the RCS and the VIRMOS-DESCART survey, allowing us for
the first time to measure both $b$ and $r$ as a function of scale.

The structure of the paper is as follows. In \S 2 we discuss the data
sets we use for our analysis. In \S3 we define the bias parameters,
and show how they can be related to observed correlation functions.
The relevant correlation functions are discussed in \S4. The actual
measurements are discussed in \S5.  The inferred bias parameters are
presented in \S6. In the appendix we demonstrate that the method
used in this paper gives accurate results if the bias paramters
vary with scale and redshift.

\vspace{0.5cm}
\section{Data}

The Red-Sequence Cluster Survey\footnote{ \tt
http://www.astro.utoronto.ca/${\tilde{\ }\!}$gladders/RCS} is a galaxy
cluster survey designed to provide a large number of clusters with
$0.1<z<1.4$ (e.g., Yee \& Gladders 2001). Here we use the $R_C$ band
data from the Northern half of the survey, which consists of 10 widely
separated patches on the sky, observed with the CFH12k camera on the
CFHT. The total area observed is 45.5~deg$^2$, but because of masking
the effective area is somewhat smaller, with a total of 42~deg$^2$
used in the lensing analysis. The data and weak lensing analysis are
described in detail in Hoekstra et al. (2002a).

The VIRMOS-DESCART survey\footnote{{\tt
http://www.astrsp-mrs.fr/virmos/} and {\tt
http://terapix.iap.fr/Descart/}} consists of four patches in 4 colors,
also observed with the CFHT12k camera. The final survey will cover 16
deg$^2$, but here we use 8.5 deg$^2$ of $I$-band data described in Van
Waerbeke et al. (2001).

These data have been used elsewhere to derive joint constraints on
$\Omega_m$ and $\sigma_8$ using the weak lensing signal caused by
large scale structure (Hoekstra et al. 2002a; Hoekstra et al. 2002b;
Pen et al. 2002; van Waerbeke et al. 2001; van Waerbeke et al. 2002).
These papers describe in detail the object detection, the shape
measurements and the corrections for the various observational
distortions (PSF anisotropy, seeing, and camera shear).  These studies
have shown that the contamination of the lensing signal by residual
systematics is small.

The accuracy of the measurement of the galaxy auto-correlation
function increases with survey area, and hence the RCS data are best
suited for this measurement. Currently, the VIRMOS-DESCART survey
provides the most accurate measurement of the matter auto-correlation
function (in particular on scales less than 10 arcminutes, where
systematics are negligible), although the RCS data also provide useful
constraints on cosmological parameters (Hoekstra et al. 2002b). On
scales less than 10 arcminutes, the accuracy of the RCS measurements
is likely limited by intrinsic alignments of the sources, whereas the
VIRMOS-DESCART measurements do not suffer from this. For the
galaxy-mass cross-correlation function we use the results from the
RCS, because the lenses used for the determination of the galaxy
auto-correlation function were selected from this survey.

Currently, we do not have redshift information for the RCS galaxies, and
we therefore select a sample of lenses on the basis of their apparent
$R_C$-band magnitude: we define a sample of lenses from the
RCS with $ 19.5<R_C<21$, which yields a total of $\sim 1.2\times10^5$ lenses.
For the measurement of the galaxy-mass cross-correlation function
we use source galaxies with $21.5<R_C<24$ which yields
$\sim 1.5\times 10^6$ sources. The weak lensing analysis of the
VIRMOS-DESCART data uses $\sim 5\times 10^5$ galaxies with $I_{AB}<24.5$.

In order to interpret the measurements, we have to know the redshift
distributions of the lens galaxies and the two populations of source
galaxies.

The CNOC2 Field Galaxy Redshift Survey (e.g., Yee et al. 2000) has
determined the redshift distribution of galaxies down to a nominal
limit of $R_C=21.5$. Hence, the results from CNOC2 provide an
excellent measure of the redshift distribution of our sample of lenses
$(19.5<R_C<21)$, for which we obtain a median redshift of $z=0.35$.

The source galaxies from the RCS and the VIRMOS-DESCART surveys have
different redshift distributions, and these galaxies are generally too
faint for spectroscopic surveys. Fortunately, photometric
redshifts work well, as demonstrated by Hoekstra et al. (2000a).  We
use photometric redshift distributions derived from both Hubble Deep
Fields (Fern{\'a}ndez-Soto et al. 1999), and
multi-colour observations done with the Very Large Telescope (Van
Waerbeke et al. 2002). We find a median redshift of
$z=0.53$ for the RCS source galaxies (with $21.5<R_C<24$).  The
galaxies from the VIRMOS-DESCART survey are fainter, and the selection
of galaxies with $I_{AB}<24.5$ gives a median redshift of $z\simeq
0.9$. Although the source redshift distributions are quite different, we
will show below that the two surveys are actually well matched, as
both probe the power spectrum at $z\sim0.35$ (see Fig.~\ref{z_probe}).

\section{Method}

To study the galaxy biasing we use a combination of the galaxy and
mass auto-correlation functions, as well as the cross-correlation
function. The statistic we use to present the results is the
``aperture mass'' $M_{\rm ap}$, which is described in detail in
Schneider et al. (1998). It is defined as (Kaiser et al. 1994)

\begin{equation}
M_{\rm ap}(\theta)=\int d^2\phi U(\phi) \kappa(\phi).
\end{equation}

Provided $U(\phi)$ is a compensated filter, i.e., $\int d\phi \phi U(\phi)=0$,
with $U(\phi)=0$ for $\phi>\theta$, the aperture mass can be
expressed in term of the observable tangential shear $\gamma_t$ using
a different filter function $Q(\phi)$ (which is a function of
$U(\phi)$),

\begin{equation}
M_{\rm ap}(\theta)=\int_0^\theta d^2\phi Q(\phi)\gamma_t(\phi).
\end{equation}

\noindent We use the filter function suggested by Schneider et al. (1998)

\begin{equation}
U(\phi)=\frac{9}{\pi \theta^2_{\rm ap}}
\left[1-\left(\frac{\phi}{\theta_{\rm ap}}\right)^2\right]
\left[\frac{1}{3}-\left(\frac{\phi}{\theta_{\rm ap}}\right)^2\right],
\end{equation}

\noindent with the corresponding $Q(\phi)$

\begin{equation}
Q(\phi)=\frac{6}{\pi\theta^2_{\rm ap}}
\left(\frac{\phi}{\theta_{\rm ap}}\right)^2
\left[1-\left(\frac{\phi}{\theta_{\rm ap}}\right)^2\right].
\end{equation}

A common definition of the ``bias'' parameter is the ratio of the
variances of the galaxy and dark matter densities, which is the
definition we will use here. In the case of deterministic, linear
biasing the galaxy density contrast $\delta_g$ is simply related to
the mass density contrast $\delta$ as $\delta_g=b\delta$ (Kaiser
1987), and the ratio of the variances is the only relevant parameter.
The galaxy number density contrast $\Delta n_g$ is then given by

\begin{equation}
\Delta n_{\rm g}(\theta)= \frac{N(\theta)-\bar N}{\bar N}=
b\int dw~p_f(w)\delta(f_K(w)\theta;w),
\end {equation}

\noindent where $\bar N$ is the average number density of lens
galaxies, $w$ is the comoving distance, $f_K(w)$ is the comoving
angular diameter distance, and where $p_f(w)dw$ corresponds to the
redshift distribution of lens galaxies. 

\noindent Then the ``aperture count'' ${\cal N}$ is given by (Schneider 1998)

\begin{equation}
{\cal N}(\theta_{\rm ap})=\int d^2\phi U(\phi)\Delta n_{\rm g}(\phi).
\end{equation}

\noindent With our choice of the filter function $U(\phi)$, we can
write the  auto-correlation of $\cal N$ as (Schneider
1998; Van Waerbeke 1998; Hoekstra et al. 2001)

\begin{equation}
\langle{\cal N}^2(\theta_{\rm ap})\rangle=2\pi b^2 \int dw h_1(w; \theta_{\rm ap}) 
\end{equation}

\noindent where $h_1(w; \theta_{\rm ap})$ is defined as 

\begin{equation}
h_1(w; \theta_{\rm ap})=\left(\frac{p_f(w)}{f_K(w)}\right)^2 P_{\rm filter}(w; \theta_{\rm ap}),
\end{equation}

\noindent and where the ``filtered'' power spectrum
$P_{\rm filter}(w;\theta_{\rm ap})$ is given by (e.g., Schneider et al. 1998)

\begin{equation}
P_{\rm filter}(w;\theta_{\rm ap})\hspace{-0.3mm}=\hspace{-1.2mm}\int\hspace{-1.0mm}
dl~l P_{\rm 3d}\hspace{-0.3mm}\left(\frac{l}{f_K(w)};w\right)
J^2(l\theta_{\rm ap}).
\end{equation}

\noindent Here $P_{\rm 3d}$ is the time-evolving 3-D power spectrum.
As has been shown by Jain \& Seljak (1997), it is important to use the
non-linear power spectrum in the calculations, and in the following we
use the results from Peacock \& Dodds (1996). The filter function
$J(\eta)$ is given by

\begin{equation}
J(\eta)=\frac{12}{\pi\eta^2}J_4(\theta)
\end{equation}

\noindent where $J_4(x)$ is the fourth order Bessel function of the first kind.

The power spectrum $P_{\rm 3d}$ contains a wealth of information about
the cosmological parameters. A measurement of $\langle {\cal
N}^2\rangle$ could be a powerful tool, provided the value of $b$ is
known. Unfortunately, the latter is not true. This is the reason why
weak lensing by large scale structure (``cosmic shear'') has become 
an important cosmological tool: it probes the (dark) matter power spectrum
directly, without having to rely on the light distribution.

\noindent The matter auto-correlation function $\langle M_{\rm
ap}\rangle^2$ is related to the power spectrum as (e.g., Schneider et
al. 1998)

\begin{equation}
\langle M_{\rm ap}^2(\theta_{\rm ap})\rangle=
\frac{9\pi}{2}\left(\frac{H_0}{c}\right)^4\Omega_m^2
\int dw h_2(w;\theta_{\rm ap}),
\end{equation}

\noindent where $\Omega_m$ is the density parameter, and  $h_2(w;\theta_{\rm ap})$ is given by

\begin{equation}
h_2(w;\theta_{\rm ap})=\left(\frac{g(w)}{a(w)}\right)^2 P_{\rm filter}(w; \theta_{\rm ap}),
\end{equation}

\noindent where $a(w)$ is the cosmic expansion factor. The function
$g(w)$ is given by

\begin{equation}
g(w)=\int_{w}^{w_H} dw' p_b(w') \frac{f_K(w'-w)}{f_K(w')},
\end{equation}

\noindent and depends on the redshift distribution of the (background)
sources $p_b(w)dw$.  It measures the ``lensing strength'' of a lens at
a distance $w$. If the lens is close to the sources, the lensing
signal decreases, whereas a large distance between the lens and the
sources results in a larger signal. Hence, $g(w)$ declines with
increasing $w$, reaching a value of 1 for $w=0$, and a value of 0 for
lenses behind the sources.

\noindent We use Eqns.~7 and~11 to relate the bias parameter $b$ to the cosmology and
measurements, and obtain

$$
b^2=\frac{9}{4}\left(\frac{H_0}{c}\right)^2
\left[\frac{\int dw h_2(w;\theta_{\rm ap})}
{\int dw h_1(w;\theta_{\rm ap})}\right]
\Omega_m^2\times
\frac{\langle {\cal N}^2(\theta_{\rm ap})\rangle}{\langle M_{\rm ap}^2(\theta_{\rm ap})\rangle}
$$

\begin{equation}
=f_1(\theta_{\rm ap},\Omega_m,\Omega_\Lambda)\times\Omega_m^2
\times\frac{\langle{\cal N}^2(\theta_{\rm ap})\rangle}
{\langle M_{\rm ap}^2(\theta_{\rm ap})\rangle}.
\end{equation}

\noindent The value of $f_1$ depends on the assumed cosmological model and
the redshift distributions of the lenses and the sources. Hence, for a 
given cosmology, the bias parameter $b$ can be determined from the observed
ratio of the galaxy and matter auto-correlation functions. Calculations
of $f_1$ as a function of aperture size show that it depends minimally
on the aperture size, and the adopted power spectrum. 

The bias relation is likely to be more complicated than the simple
case of linear deterministic biasing: the actual relation depends on
the process of galaxy formation, and might be stochastic, non-linear
or both. Hence, there is no reason that $b$ is constant with scale,
but as we demonstrate in Appendix~A, we can still use Eqn.~12 as long
as $b$ varies slowly with scale. Another complication is that $b$ is
likely to depend on redshift, and conseqently the derived value for
$b$ is a redshift averaged value for the lenses in our sample.

If the bias parameter depends on redshift and scale, it is important
that both the galaxy and matter auto-correlation functions probe the
power spectrum at the same ``effective'' redshift. In addition, the
interpretation is facilitated if the measurements probe a relatively
small range in redshift. To examine this in more detail, it is useful
to plot the $h_1(w;\theta_{\rm ap})$ and $h_2(w;\theta_{\rm ap})$ as
a function of $w$.

The results are presented in Figure~\ref{z_probe}. The integrands have
been normalized to a peak value of unity. The thick solid line shows
$h_1(w;\theta_{\rm ap})$ as a function of $w$ for a fiducial
$\theta_{\rm ap}=5'$. We used the observed redshift distribution of
galaxies with $19.5<R_C<21$ from the CNOC2 survey, which gives rise to
the wiggles in the function. This result indicates that we probe the
power spectrum at an effective redshift $\sim 0.35$. The thick dashed
line shows $h_2(w;\theta_{\rm ap})$, for the VIRMOS-DESCART
survey. This function is rather broad, which reflects the fact that
weak lensing probes the power spectrum over a relatively large
redshift range. However, comparison with the integrand for the galaxy
auto-correlation function shows that both peak at the same
redshift. Hence, we can use the ratio of $\langle {\cal N}^2\rangle$
and $\langle M_{\rm ap}^2\rangle$ measured at the same aperture size
$\theta_{\rm ap}$. 

The lensing is maximal when the the lens is half-way between the
source and the observer. Hence, the weight function $h_2(w)$ reaches
its peak value approximately half-way between the observer and the
source. Compared to the background galaxies from the VIRMOS-DESCART
survey, the RCS sources are at lower redshift, and consequently the
latter will probe the power spectrum at a lower redshift. This is
indicated by the thin dashed line in Figure~\ref{z_probe}, which
shows $h_2(w;\theta_{\rm ap})$ using the RCS background galaxy
redshift distribution. It is a poor match to the galaxy auto-correlation
function.  We therefore use the mass auto-correlation function
measured from the VIRMOS-DESCART survey.

So far, we have only used the auto-correlation functions. We can,
however, also measure the galaxy-mass cross-correlation function
$\langle{\cal N}M_{\rm ap}\rangle$, which can be used to quantify how
well the mass distribution correlates with the light distribution.

The galaxy-mass cross-correlation function $\langle{\cal N}M_{\rm
ap}\rangle$ is related to the power spectrum as (Schneider 1998; Van
Waerbeke 1998; Hoekstra et al. 2001)

\begin{equation}
\langle M_{\rm ap}(\theta_{\rm ap}){\cal N}(\theta_{\rm ap})\rangle  
= 3\pi \left(\frac{H_0}{c}\right)^2 \Omega_m b r
\int dw h_3(w;\theta_{\rm ap}),
\end{equation}

\noindent where $h_3(w;\theta_{\rm ap})$ is defined as 

\vbox{
\vspace{-0.3cm}
\begin{center}
\leavevmode 
\hbox{% 
\epsfxsize=\hsize 
\epsffile[40 160 560 710]{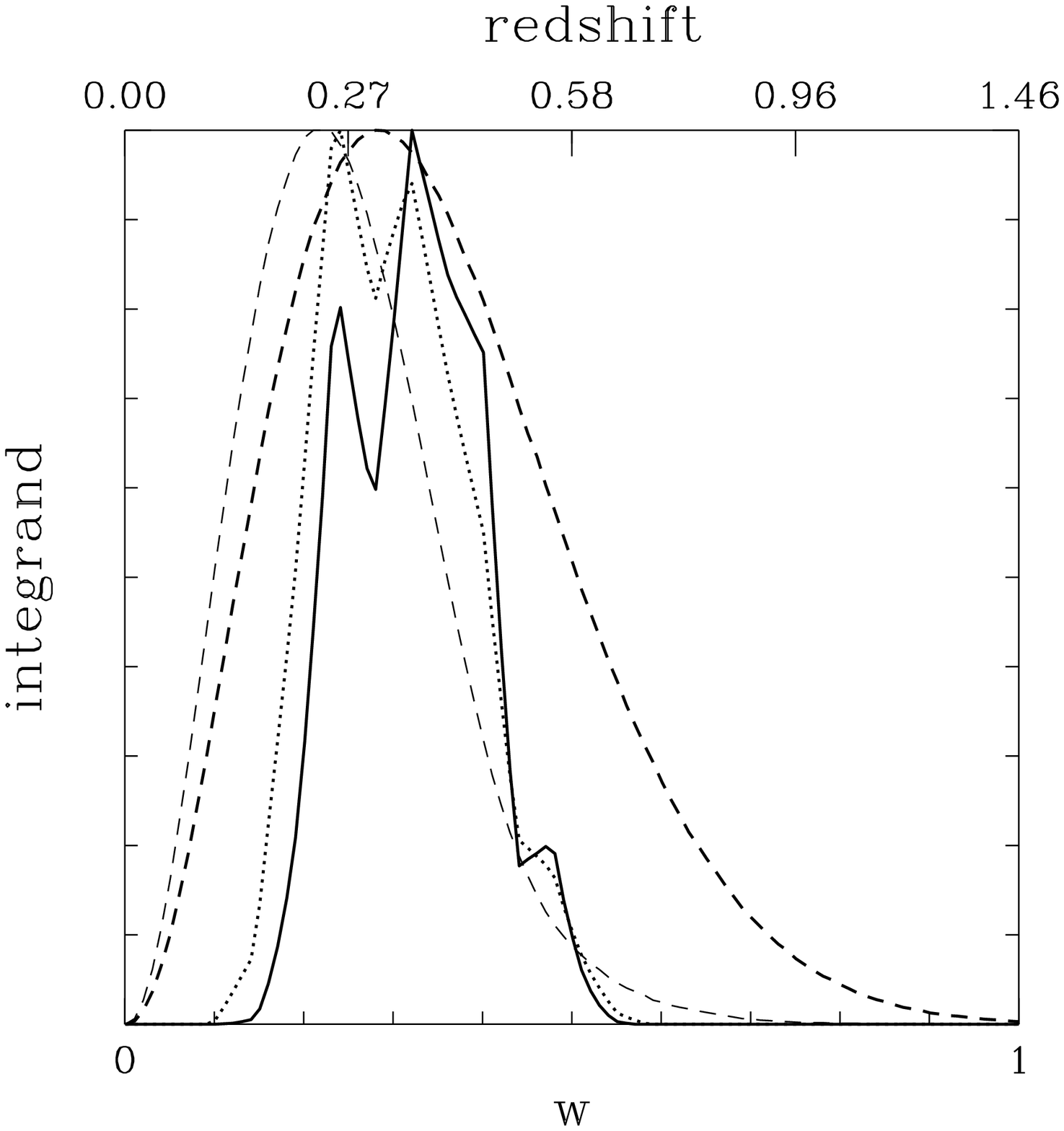}}
\figcaption{\footnotesize The different integrands of the integrals in
Eqns.~14 and~17 for a fiducial aperture size $\theta_{\rm ap}=5'$.
The integrands have been normalized to have the same maximum
value. The thick solid line shows $h_1(w)$ as a function of $w$. The
upper axis indicates the corresponding redshift.  We used the observed
redshift distribution of galaxies with $19.5<R_C<21$ from the CNOC2
survey, which gives rise to the wiggles in the function. This result
indicates that we probe the power spectrum (and the bias parameters)
at an effective redshift $\sim 0.35$. The thick dashed line shows
$h_2(w)$, for the VIRMOS-DESCART survey. This function is rather
broad, which reflects the fact that weak lensing probes the power
spectrum over a relatively large redshift range. However, comparison
with the integrand for the galaxy auto-correlation function shows that
both peak at the same redshift. Hence, we can use the ratio of
$\langle {\cal N}^2\rangle$ and $\langle M_{\rm ap}^2\rangle$ measured
at the same aperture size $\theta_{\rm ap}$. The same integrand using
the RCS background galaxy redshift distribution gives the thin dashed
line, which is a poor match to galaxy auto-correlation function. The
integrand for the galaxy-mass cross-correlation function, $h_3(w)$, is
indicated by the thick dotted line, which overlaps well with the
galaxy auto-correlation function.
\label{z_probe}}
\end{center}}

\begin{equation}
h_3(w;\theta_{\rm ap})=\frac{p_f(w) g(w)}{a(w) f_K(w)} P_{\rm filter}(w; \theta_{\rm ap}).
\end{equation}

\noindent The parameter $r$ in Eqn.~15 is the galaxy-mass
cross-correlation coefficient (e.g., Pen 1998; Dekel \& Lahav 1999;
Somerville et al. 2001), which is a measure of non-linear stochastic
biasing. In the case of deterministic linear biasing, $r=1$.

The redshift distributions of the lenses and the sources used to
measure the galaxy-mass cross-correlation function partly overlap. A
source in front of a lens contributes no signal, and therefore lowers
the lensing signal. The overlap of the redshift distributions is
naturally taken into account by Eqn.~15. 

To prove this statement, we first consider a sheet of lenses at a
distance $\omega_f$. The ``lensing strength'' for these lenses is
given by $g(w_f)$, to which only background galaxies with
$\omega>\omega_f$ contribute. The larger the $\omega_f$, the lower
both the observed and predicted lensing signals will be (and they are
lower by the same factor).  Hence, Eqn.~15 gives the correct result
for a sheet of lenses at a given redshift. The actual redshift
distribution of lenses can be considered a superposition of many
sheets at different redshifts, and consequently Eqn.~15 is correct for
any combination of lens and source redshift distributions.

\noindent The combination of Eqns.~7, 11, and 15 relates the value of
$r$ to the observed correlation functions and the cosmology, and is
given by

$$
r=\frac{\sqrt{\int dw h_1(w;\theta_{\rm ap})}
\sqrt{\int dw h_2(w;\theta_{\rm ap})}}
{\int dw h_3(w;\theta_{\rm ap})}
\times \frac{\langle M_{\rm ap}{\cal N}\rangle}
{\langle M_{\rm ap}^2\rangle^{1/2}\langle {\cal N}^2\rangle^{1/2}}
$$

\begin{equation}
=f_2(\theta_{\rm ap},\Omega_m,\Omega_\Lambda)\times
\frac{\langle M_{\rm ap}(\theta_{\rm ap})\cal{N}(\theta_{\rm ap})\rangle}
{\sqrt{\langle {\cal N}^2(\theta_{\rm ap})\rangle
\langle M_{\rm ap}^2(\theta_{\rm ap})\rangle}}.
\end{equation}

\noindent Similar to $f_1$, the value of $f_2$ depends minimally on
$\theta_{\rm ap}$ and the assumed power spectrum (Van Waerbeke 1998).

The integrand $h_3(w;\theta_{\rm ap})$ of the integral in Eqn.~15 is
also shown in Figure~\ref{z_probe}, and it matches the results for the
galaxy auto-correlation function well.  As for the bias parameter $b$,
the value of $r$ can depend on scale and redshift. However, because of
the properties of the aperture statistics, described in Appendix~A, we
can measure the scale dependence of $r$ directly from Eqn.~17. As for
$b$, the inferred value for $r$ is a redshift averaged value for the
lenses in our sample.

Our definitions of $b$ and $r$ are chosen such that they can be
related directly to the observed correlation functions. However, both
$b$ and the cross-correlation coefficient $r$ mix non-linear and
stochastic effects (Dekel \& Lahav 1999; Somerville et al. 2001),
which we currently cannot disentangle with weak lensing.

Dekel \& Lahav (1999) defined a combination of 3 parameters ($\tilde
b$, $\hat b$, and $\sigma_b$), which separate non-linear and
stochastic biasing. They parametrize the stochasticity by the
parameter $\sigma_b$, the local biasing scatter. The non-linear
biasing is characterized by the ratio $\tilde b/\hat b$, where $\hat
b$ is the slope of the linear regression of $\delta_g$ on $\delta$
(i.e., the natural generalization of the linear biasing parameter).

From lensing we can only measure combinations of these parameters.  We
can, however, relate the observables $b$ and $r$ to the the parameters
from Dekel \& Lahav (1999) in the two limiting cases.  If we assume
that the bias is purely non-linear and deterministic (i.e.,
$\sigma_b=0$), we obtain $\tilde b=b$, and $\tilde b/\hat
b=1/r$. Hence, the inverse of the cross-correlation coefficient
measures the amount of non-linear biasing. Likewise, in the case of
linear and stochastic biasing (i.e., $\tilde b=\hat b$) we obtain
$\sigma_b=b\sqrt{1-r^2}$.

\section{Determination of the correlation functions}

A straightforward implementation of the method described above is to
tile the survey area with apertures, and compute $\langle M_{\rm
ap}{\cal N}\rangle(\theta)$ and $\langle{\cal N}^2\rangle(\theta)$
directly from the data.  To do so, one can use the estimators
(Schneider 1998):

$$
\tilde M_{\rm ap}= \pi\theta_{\rm ap}^2
\frac{\sum_{i=1}^{N_b} Q(\theta_i) w_i \gamma_{{\rm t},i}}
{\sum_{i=1}^{N_b} w_i},~{\rm and}~
\tilde {\cal N}=\frac{1}{\bar N} \sum_{i=1}^{N_f} U(\theta_i),
$$

\noindent where $N_f$ and $N_b$ are respectively the number of lens
and source galaxies found in the aperture of radius $\theta_{\rm ap}$,
$\bar N$ is the average number density of lenses, and $\gamma_{t,i}$ is
the observed tangential shear of the $i$th background galaxy. The weights
$w_i$ correspond to the inverse square of the uncertainty in the shear
measurement (Hoekstra et al. 2000).

However, this procedure has the disadvantage that it assumes a
contiguous data set; i.e., there are no holes in the data. In real
data, regions that are contaminated by bright stars, bad columns, etc.,
have to be masked.  Although the masking of the RCS data is not
severe, we will use a different approach, which is much less sensitive
to the geometry of the survey.

Instead we measure the ensemble averaged tangential shear as a
function of radius around the sample of lenses (``galaxy-galaxy
lensing'' signal) and use these results to derive $\langle M_{\rm
ap}{\cal N}\rangle$. We use the angular two-point correlation function
to estimate $\langle {\cal N}^2\rangle$. The mass auto-correlation
function $\langle M_{\rm ap}^2\rangle$ is derived from the observed
ellipticity correlation functions (e.g., Pen et al. 2002; van Waerbeke
et al. 2002; Hoekstra et al. 2002b).  We show below how these observed
correlation functions, which do not require contiguous survey areas,
can be easily related to the aperture mass correlation functions.

We first consider the angular two-point correlation function
$\omega(\theta)$, which is related to the power spectrum through

$$
\omega(\theta) =  \langle \Delta n_g(0) \Delta n_g (\theta)\rangle=
$$
\begin{equation}
\frac{b^2}{2\pi}\int dw \left(\frac{p_f(w)}{f_K(w)}\right)^2
\int dl~lP_{\rm 3d}\hspace{-0.3mm}\left(\frac{l}{f_K(w)};w\right) 
{\rm J}_0(l\theta),
\end{equation}

\noindent where ${\rm J}_0(x)$ is the zeroth order Bessel function of
the first kind.  We can define the effective (projected) power
spectrum of the angular correlation function $P_\omega(l)$ as

\begin{equation}
P_\omega(l)=  b^2\int dw \left(\frac{p_f(w)}{f_K(w)}\right)^2
P_{\rm 3d}\hspace{-0.3mm}\left(\frac{l}{f_K(w)};w\right).
\end{equation}

\noindent Thus we obtain for the angular correlation function

\begin{equation}
\omega(\theta)=\frac{1}{2\pi}\int dl~l P_\omega(l) {\rm J}_0(l\theta).
\end{equation}

\noindent For the auto-correlation function of ${\cal N}$ we have
(see Eqns.~7 and 9)

\begin{equation}
\langle{\cal N}^2\rangle(\theta)=2\pi \int dl~l P_\omega(l)
\left[\frac{12 J_4(l \theta)}
{\pi (l \theta)^2}\right]^2.
\end{equation}

Schneider, van Waerbeke \& Mellier (2002) have shown that it is
rather straightforward to transform one correlation function into
another. Using the orthogonality of Bessel functions, we obtain

\begin{equation}
P_\omega(l)=2\pi \int d\vt~\vt \omega(\vt) {\rm J}_0(l\vt),
\end{equation}

\noindent which we can use to relate $\langle{\cal N}^2\rangle(\theta)$
to $\omega(\theta)$. Doing so, we obtain

\begin{equation}
\langle{\cal N}^2\rangle(\theta)=\int d\vt~\frac{\vt}{\theta^2}
\omega(\vt) T_+\left(\frac{\vt}{\theta}\right),
\end{equation}

\noindent where the function $T_+(x)$ is defined as (Schneider
et al. 2002)

\begin{equation}
T_+(x)=576\int_0^\infty \frac{dt}{t^3} {\rm J}_0(xt)
\left[{\rm J}_4(t)\right]^2.
\end{equation}

\noindent Schneider et al. (2002) found an expression for $T_+(x)$ in terms
of elementary functions

\begin{eqnarray}
&&\hspace{-8mm}T_+(x)=\frac{6(2-15x^2)}{5}
\left[1-\frac{2}{\pi}\arcsin(x/2)\right]\nonumber\\
&&\hspace{-8mm}+\frac{x\sqrt{4-x^2}}{100\pi}(120+2320x^2-754x^4+132x^6-9x^8),
\end{eqnarray}

\noindent for $x\le 2$. $T_+(x)$ vanishes for $x>2$. Therefore the
integral in Eqn.~19 extends only over $0\le\vt\le 2\theta$. Hence we
need to measure the angular correlation function out to twice the
aperture size in order to compute $\langle{\cal
N}^2\rangle(\theta)$. 

We now turn to the measurement of the galaxy-mass cross-correlation
function. In this case the azimuthally averaged tangential shear
around the lens galaxies is a very useful statistic (e.g., Fischer et
al. 2000). To derive the galaxy-mass correlation function we use the
relation between the average convergence $\bar\kappa(\theta)$ inside a
circular aperture of radius $\theta$ and the mean tangential shear
along the boundary of the aperture

\begin{equation}
\langle \gamma_t\rangle(\theta)=
-\frac{1}{2}\frac{d\bar\kappa(\theta)}{d\ln\theta}.
\end{equation}

The galaxy-mass correlation function is defined as $\langle \Delta
n_g(0) \gamma_t(\theta)\rangle= \langle \gamma_t\rangle(\theta)$, and
is related to the power spectrum as (Kaiser 1992; Guzik \& Seljak
2001)

\begin{eqnarray}
\langle\gamma_t\rangle(\theta)=&&\frac{3\Omega_m}{4\pi}\left(\frac{H_0}{c}\right)^2 br
\int dw \frac{g(w) p_f(w)}{a(w)f_K(w)}\times\nonumber\\
&&\int dl~l P_{\rm 3d}\hspace{-0.3mm}\left(\frac{l}{f_K(w)};w\right) 
{\rm J}_2(l\theta),
\end{eqnarray}

\noindent where ${\rm J}_2(x)$ is the second order Bessel function of
the first kind. We assume that the cross power spectrum is related to
the matter power spectrum by $brP_{\rm 3d}(k)$. As discussed above,
the assumption of a constant value of $b$ and $r$ does not change the
interpretation of our measurements.

We define the effective (projected) power spectrum for the
galaxy-mass correlation function $P_\gamma(l)$ as

\begin{eqnarray}
P_\gamma(l)=&& \frac{3}{2}\left(\frac{H_0}{c}\right)^2\Omega_m b r
\times\nonumber\\ &&\int dw \frac{g(w) p_f(w)}{a(w)f_K(w)}
P_{\rm 3d}\hspace{-0.3mm}\left(\frac{l}{f_K(w)};w\right),
\end{eqnarray}

\noindent which gives

\begin{equation}
\langle\gamma_t\rangle(\theta)=\frac{1}{2\pi}
\int dl~l P_\gamma(l) {\rm J}_2(l\theta),
\end{equation}

\noindent and

\begin{equation}
\langle{\cal N}M_{\rm ap}\rangle(\theta)=2\pi
\int dl~l P_\gamma(l) \left[\frac{12 J_4(l \theta)}
{\pi (l \theta)^2}\right]^2
\end{equation}

\noindent As before we can express $P_\gamma(l)$ by an integral
over the galaxy-mass cross-correlation function, and we obtain

\begin{equation}
\langle{\cal N}M_{\rm ap}\rangle(\theta)=
\int d\vt~\frac{\vt}{\theta^2}
\langle\gamma_t\rangle(\vt) F\left(\frac{\vt}{\theta}\right),
\end{equation}

\noindent where we define the function $F(x)$ as

\begin{equation}
F(x)=576\int_0^\infty \frac{dt}{t^3} {\rm J}_2(xt)
\left[{\rm J}_4(t)\right]^2.
\end{equation}

We did not find a simple expression for this function, but it is well
behaved for $0\le x\le 2$, and vanishes for $x>2$. Hence, as before,
the integral in Eqn.~31 extends only over $0\le\vt\le 2\theta$.

The mass auto-correlation function $\langle M_{\rm ap}^2\rangle$ can
be obtained by measuring the ellipticity correlation functions, which
are given by

\begin{equation}
\xi_{\rm tt}(\theta)=\frac{\sum_{i,j}^{N_s} w_i w_j 
\gamma_{{\rm t},i}({{\bf x}_i}) \cdot \gamma_{{\rm t},j}({{\bf x}_j})}
{\sum_{i,j}^{N_b} w_i w_j},
\end{equation}

\begin{equation}
\xi_{\rm rr}(\theta)=\frac{\sum_{i,j}^{N_s} w_i w_j 
\gamma_{{\rm r},i}({{\bf x}_i}) \cdot \gamma_{{\rm r},j}({{\bf x}_j})}
{\sum_{i,j}^{N_b} w_i w_j},
\end{equation}

\noindent where $\theta=|{\bf x}_i-{\bf x}_j|$. $\gamma_{\rm t}$ and
$\gamma_{\rm r}$ are the tangential and 45 degree rotated shear in the
frame defined by the line connecting the pair of galaxies. For the
following, it is more useful to consider

\begin{equation}
\xi_+(\theta)=\xi_{\rm tt}(\theta)+\xi_{\rm rr}(\theta),{\rm~and~}
\xi_-(\theta)=\xi_{\rm tt}(\theta)-\xi_{\rm rr}(\theta),
\end{equation}

\noindent i.e., the sum and the difference of the two observed
correlation functions. As shown by Crittenden et al. (2002), one can
derive ``E'' and ``B''-mode correlation functions by integrating
$\xi_+(\theta)$ and $\xi_-(\theta)$ with an appropriate window
function (see Pen et al. 2002 for an application to the VIRMOS-DESCART
data).

The ``E'' and ``B''-mode aperture masses are computed from the
ellipticity correlation functions using (e.g., Crittenden et al. 2002;
Schneider et al. 2002)

\begin{equation}
\langle M_{\rm ap}^2\rangle(\theta)=\int \frac{d\vt~\vt}
{2 \theta_{\rm ap}^2}
\left[\xi_+(\vt) T_+\left(\frac{\vt}{\theta_{\rm ap}}\right)
+\xi_-(\vt) T_-\left(\frac{\vt}{\theta_{\rm ap}}\right)\right],
\end{equation}

\noindent and

\begin{equation}
\langle M_\perp^2\rangle(\theta)=\int \frac{d\vt~\vt}
{2\theta_{\rm ap}^2}
\left[\xi_+(\vt) T_+\left(\frac{\vt}{\theta_{\rm ap}}\right)
-\xi_-(\vt) T_-\left(\frac{\vt}{\theta_{\rm ap}}\right)\right].
\end{equation}

\noindent The expression for $T_+(x)$ is given by Eqn.~22, whereas $T_-(x)$ is given by

\begin{equation}
T_-(x)=\frac{192}{35\pi} x^3\left(1-\frac{x^2}{4}\right)^{7/2},
\end{equation}

\noindent for $x\le 2$, and $T_-(x)$ vanishes for $x>2$.

The ``B''-mode aperture mass $\langle M_\perp^2\rangle$ provides a
quantitative estimate of the systematics, since gravitational lensing
only produces an ``E''-mode. Residual systematics (e.g., imperfect
correction for the PSF anisotropy) or intrinsic alignments will give
rise to a ``B''-mode. A similar test exists for the galaxy-mass
cross-correlation function: in the absence of systematics, the average
signal around the lenses, when the sources are rotated by 45 degrees,
should vanish.

\vbox{
\vspace{-0.3cm}
\begin{center}
\leavevmode 
\hbox{% 
\epsfxsize=\hsize 
\epsffile[40 190 560 710]{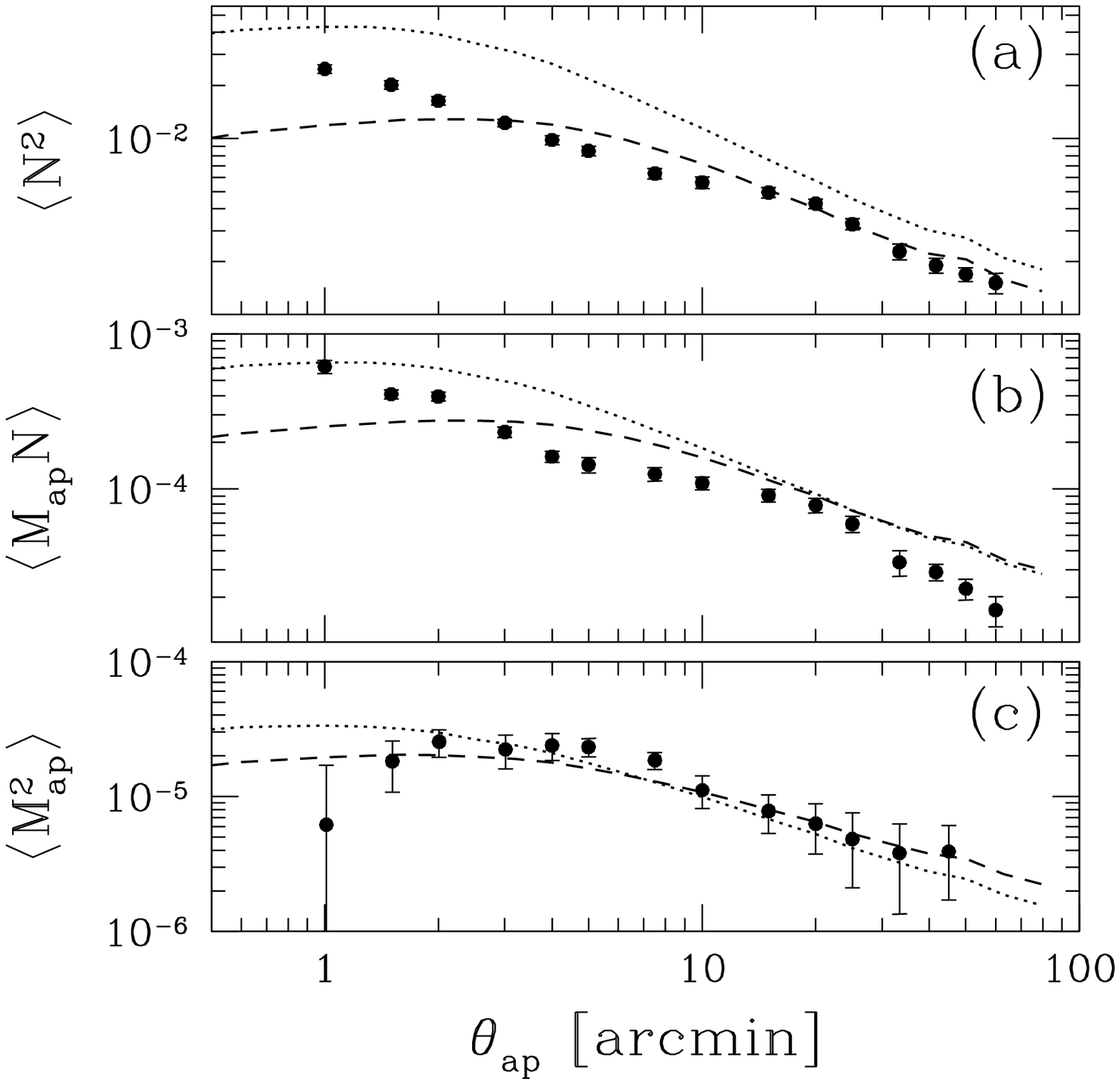}}
\figcaption{\footnotesize The measurements of $\langle {\cal
N}^2\rangle$ (panel~a), and $\langle {\cal N}M_{\rm ap}\rangle$
(panel~b) as a function of angular scale from the RCS data. The
largest scale corresponds to an aperture radius of 1 degree. Panel~c
shows $\langle M_{\rm ap}^2\rangle$ as a function of angular scale
from the VIRMOS-DESCART data (largest scale is 45 arcminutes). The
error bars for $\langle M_{\rm ap}^2\rangle$ have been increased
to account for the unknown correction for the observed ``B''-mode.
For reference (they are not fitted to the measurements) we have also
plotted model predictions (for $b=1$ and $r=1$) for an OCDM cosmology
(dotted line; $\Omega_m=0.3$, $\Omega_\Lambda=0$, $\sigma_8=0.9$, and
$\Gamma=0.21$) and a $\Lambda$CDM cosmology (dashed line;
$\Omega_m=0.3$, $\Omega_\Lambda=0.7$, $\sigma_8=0.9$, and
$\Gamma=0.21$).  Note that the points at different scales are slightly
correlated.
\label{corrfn}}
\end{center}}

\section{Measurements}

In this section we present the measurements of the galaxy auto-correlation
and the galaxy-mass cross-correlation function 
$\langle M_{\rm ap}(\theta_{\rm ap}){\cal N}(\theta_{\rm ap})\rangle$,
which were obtained from the RCS data. The  mass auto-correlation
function $\langle M_{\rm ap}^2(\theta_{\rm ap})\rangle$ was measured
by van Waerbeke et al. (2002) from the VIRMOS-DESCART survey.

As described in \S4, we do not measure $\langle{\cal N}^2\rangle$
directly from the data. Instead we measure the angular correlation
function $\omega(\theta)$ and use Eqn.~23 to determine $\langle{\cal
N}^2\rangle$. To measure $\omega(\theta)$, we use the well known
estimator (Landy \& Szalay 1993)

\begin{equation}
\omega(\theta)=\frac{DD-2DR+RR}{RR},
\end{equation}

\noindent where $DD$, $DR$, and $RR$ are pair counts in bins of
$\theta+\delta\theta$ of the data-data, data-random, and random-random
points respectively. For each patch, we create 24 mock catalogs by
placing the lens galaxies at random positions in the unmasked regions
of the data. We note our sample of lenses is complete within the unmasked 
regions: the galaxies are sufficiently faint that they are not
saturated, and they are several magnitudes brighter than the
detection limit. Furthermore, the uncertainties in the photometry are
small ($<3\%$, Gladders et al. in preparation), and do not affect the
measurement of the angular correlation function of the lens galaxies.

In order to determine the true angular correlation function one needs
to apply an integral constraint correction to the observed correlation
function. The determination of this correction is not trivial, and can
introduce a significant systematic uncertainty.  However, since the
integral $\int dx~x T_+(x)$ vanishes, $\langle{\cal N}^2\rangle$ has
the nice property that it is independent of the integral constraint
correction. In addition, the measurements on different scales are only
mildly correlated.

Although we use the observed angular correlation function as an
intermediate step in our analysis, it is useful to compare the results
to previous work. The observed angular correlation function is well
approximated by a power law with slope -0.7, and an amplitude
$\omega(1')=0.115\pm0.005$. Unfortunately, a direct
comparison with the literature is difficult, because of the different
sample selection (magnitude limits, filters). We can, however, make a
crude comparison with the results from Postman et al. (1998), who
measured the angular correlation function in the $I$-band.  The mean
$R_C$-band magnitude of our sample is $\sim 20.5$, which corresponds
to a mean $I$-band magnitude of $\sim 19.8$ (based on the colors of
galaxies observed in the CNOC2 survey). Postman et al. (1998) list a
value of $\omega(1')=0.136\pm0.021$ for galaxies with a
median magnitude of $I=19.6$, which is in reasonable agreement with our
result.

We measured $\langle{\cal N}^2\rangle$ for each of the 10 RCS patches,
and used the scatter in these measurements to estimate the error bars
on the galaxy auto-correlation function. The errors therefore also
include cosmic variance. The results, measured out to 1 degree, are
presented in Figure~\ref{corrfn}a.

\vbox{
\vspace{-0.3cm}
\begin{center}
\leavevmode 
\hbox{% 
\epsfxsize=7.3cm 
\epsffile[60 160 600 730]{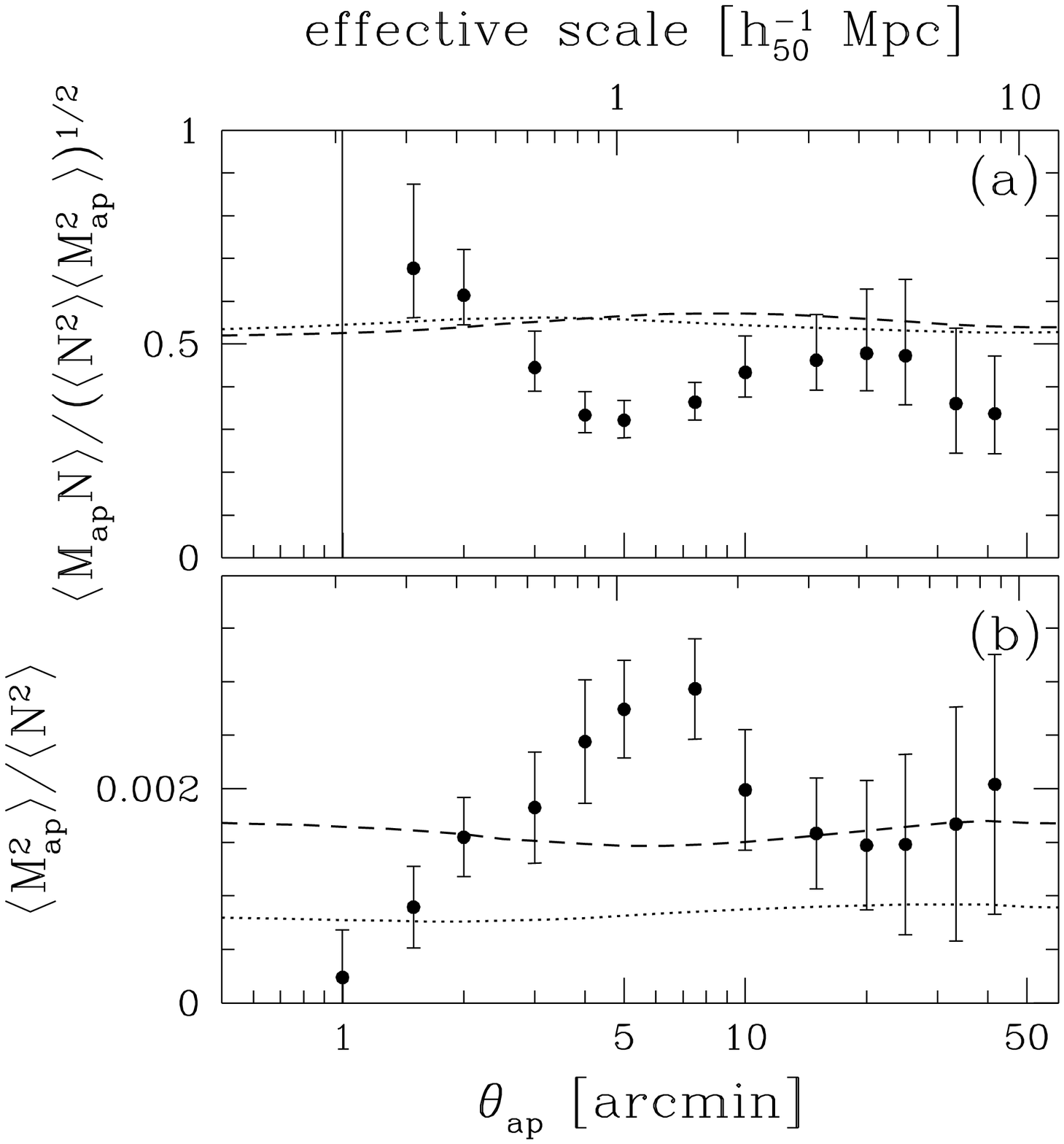}} 
\figcaption{\footnotesize (a) The observed ratio $\langle M_{\rm
ap}{\cal N}\rangle/(\langle {\cal N}^2\rangle\langle M_{\rm
ap}^2\rangle)^{1/2}$ as a function of aperture size. (b) The ratio of
the mass auto-correlation $\langle M_{\rm ap}^2\rangle$ and the galaxy
auto-correlation $\langle {\cal N}^2\rangle$ as a function of aperture
size. The measurements at different scale are slightly correlated, and
the error bars correspond to the 68\% confidence intervals.
The upper axis indicates the effective physical scale probed by the
compensated filter $U(\phi)$ at the median redshift of the lenses
$(z=0.35)$. The dotted lines are the predictions of an OCDM model,
whereas the dashed lines correspond to the $\Lambda$CDM model for
$b=1$, and $r=1$. In both cases, if $b$, and $r$ are constant with
scale, we expect to observe ratios that are virtually constant.  The
observed ratios show significant variation with scale, thus implying
that both $b$ and $r$ vary.
\label{ratios}}
\vspace{-0.5cm}
\end{center}}

\vbox{
\vspace{-0.3cm}
\begin{center}
\leavevmode 
\hbox{% 
\epsfxsize=8.3cm 
\epsffile[40 140 600 730]{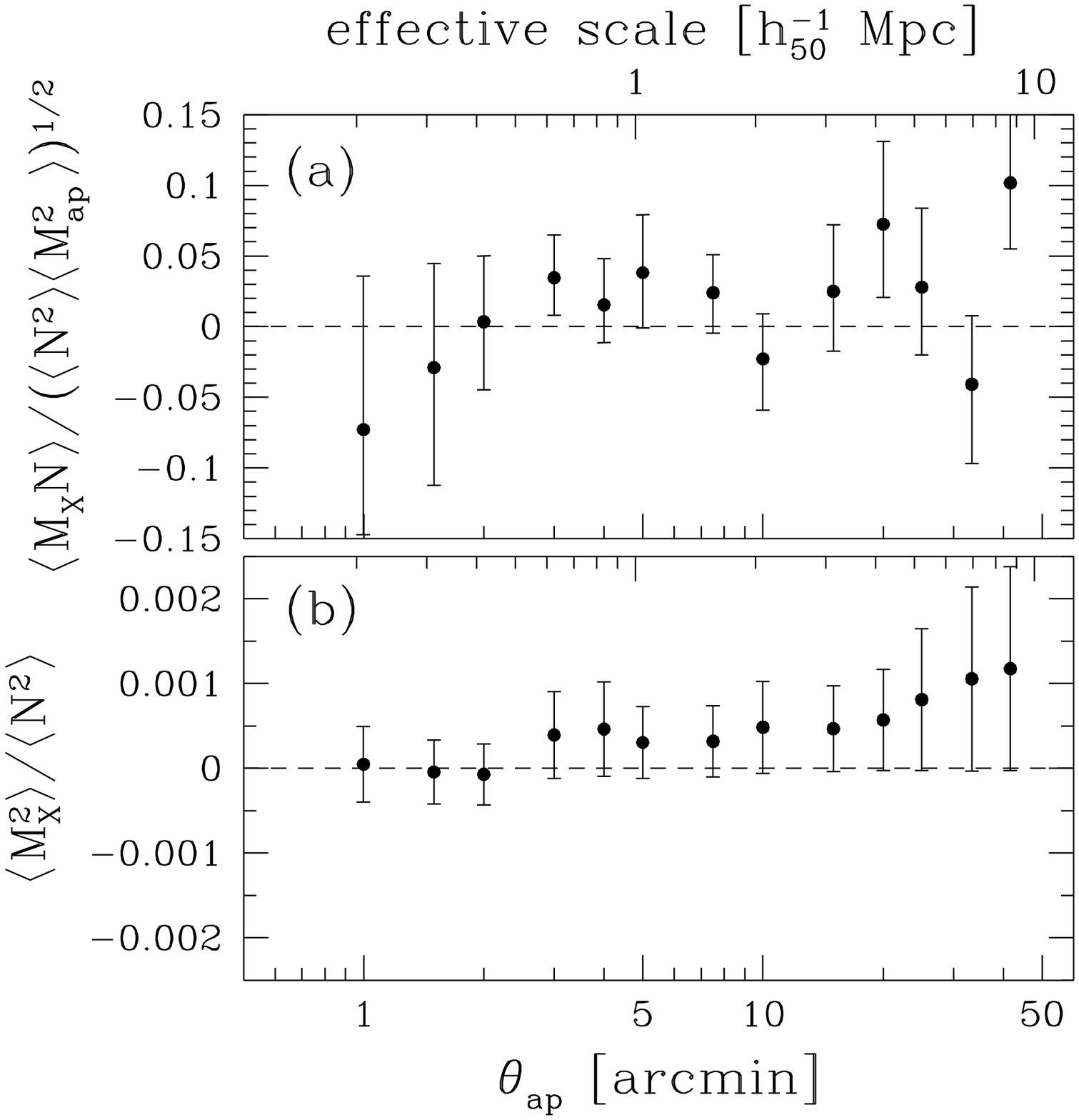}} 
\figcaption{\footnotesize (a) The observed ratio $\langle M_{\rm
X}{\cal N}\rangle/(\langle {\cal N}^2\rangle\langle M_{\rm
ap}^2\rangle)^{1/2}$ as a function of aperture size. This corresponds
to the results when the phase of the shear is increased by $\pi/2$.
If the signal presented in Figure~\ref{ratios}a is caused by
lensing it should vanish here, as it does. (b) The ratio of
the ``B''-mode auto-correlation $\langle M_{\rm X}^2\rangle$ and the galaxy
auto-correlation $\langle {\cal N}^2\rangle$ as a function of aperture
size. The results suggest a residual ``B''-mode, which is attributed
to systematics (note that the points at different scales are
somewhat correlated). The amplitude of this signal is added to the error
estimates in Figure~\ref{ratios}b as a conservative limit. We note
that the amplitude of the ``B''-mode signal is small compared to
the lensing (``E''-mode) signal.
\label{ratio_x}}
\end{center}}

Figure~\ref{corrfn}b shows the cross-correlation function $\langle
M_{\rm ap}{\cal N}\rangle$ as measured from the RCS.  To derive the
cross-correlation function we measured the azimuthally averaged
tangential shear around the lens galaxies in bins of 1 arcsecond.
We then used Eqn.~31 to relate the observed tangential shear
profile to $\langle M_{\rm ap}{\cal N}\rangle$. As before, we
measured the cross-correlation for the 10 RCS patches, and used
the scatter in the measurements to estimate the errorbars.

The mass auto-correlation function $\langle M_{\rm ap}^2\rangle$,
presented in Figure~\ref{corrfn}c, was taken from van Waerbeke et
al. (2002). It was derived from the observed ellipticity correlation
functions (see Eqn.~36). The largest scale measurement of $\langle
M_{\rm ap}^2\rangle$ from the VIRMOS-DESCART survey is 45
arcminutes. The error bars in Figure~\ref{corrfn}c have been increased
to account for the unknown correction for the observed ``B''-mode 
(van Waerbeke et al. 2002).

The shapes and amplitudes of the correlation functions presented in
Figure~\ref{corrfn} depend on the power spectrum. As discussed in \S3,
we can remove the dependence on the power spectrum by taking ratios of
the correlation functions. Since the various correlation functions
probe the power spectrum at the same redshift (see
Fig.~\ref{z_probe}), we can take the ratios of measurements at the
same angular scale. If the surveys were not well matched, we would
have had to compare the measurements at different angular scales, in
order to ensure that we probe the power spectrum at the same physical
scale. In addition, one has to take into account the evolution of the
power spectrum in such a situation.

The observed ratio $\langle M_{\rm ap}{\cal N}\rangle/(\langle {\cal
N}^2\rangle\langle M_{\rm ap}^2\rangle)^{1/2}$ as a function of
aperture size is presented in Figure~\ref{ratios}a.
Figure~\ref{ratios}b gives the ratio of $\langle M_{\rm ap}^2\rangle$
and $\langle {\cal N}^2\rangle$.  The errorbars on the ratios
correspond to the 68\% confidence limits, and have been determined
from a Monte Carlo simulation using the uncertainties in the
measurements of the observed correlation functions (which were assumed
to be Gaussian).  For reference, we have also indicated the effective
physical scale ($\sim$ FWHM of the filter function) probed by the
compensated filter $U(\phi)$ corresponding to a redshift of $z=0.35$.

For reference, Figure~\ref{ratios} also shows the ratios for an OCDM
and a $\Lambda$CDM model in the case $b=1$ and $r=1$ (i.e., we have
plotted $\Omega_m^2\times f_1$, and $f_2$). These model values are
virtually constant with scale if $b$ and $r$ are constant. Also note
that the value for the cross-correlation is almost the same for both
cosmologies, whereas the ratio of the auto-correlation functions
differs by almost a factor~2.

To examine possible systematic effects, we also computed the results
when the sources are rotated by $45^\circ$. This signal should vanish
in the case of lensing. The results of this powerful test for the
cross-correlation are presented in Figure~\ref{ratio_x}, and are
indeed consistent with no signal. The results for the auto-correlation
do show some residual systematics. The amplitude of this signal is
added to the error estimates in Figure~\ref{ratios}b as a conservative
limit (van Waerbeke et al. 2002). We note that the amplitude of the
``B''-mode signal is small compared to the lensing (``E''-mode)
signal. Based on these results we conclude that the accuracy of our
measurements (in particular the galaxy-mass cross-correlation) is not
limited by systematics.

\section{Discussion}

The results presented in Figure~\ref{ratios} suggest significant
variation of both $b$ and $r$ with scale. We convert the
observed ratios into estimates for the bias parameters for
the currently favored cosmological model $(\Omega_m=0.3,~\Omega_\Lambda=0.7)$,
using Eqns.~14 and 17.

Figure~\ref{bias}a shows the inferred value of the galaxy-mass
cross-correlation coefficient $r$ as a function of scale. Because of
the small value of $\langle M_{\rm ap}^2\rangle$ at 1~arcminute,
the uncertainties in both $b$ and $r$ are very large, and we
have omitted this point.

We find that on small scales the measurements are consistent with a
value $r\sim1$. This good correlation between mass and light on small
scales (i.e., around galaxies) indicates that (luminous) galaxies are
surrounded by massive halos. We note that with our definition $r$ can
be larger than 1. On the largest scales the results are consistent
with $r=1$. On scales $\sim 1h_{50}^{-1}$ Mpc $r$ is significantly
lower than unity, with a minimum value of $r=0.57^{+0.08}_{-0.07}$
(68\% confidence).

Figure~\ref{bias}b shows that the scale dependence of $b$ is very
similar to that of $r$ (also see Figure~\ref{b_over_r}).  We find that
$b$ is smaller than unity on scales $\sim 1-2 h_{50}^{-1}$ Mpc, with a
minimum value of $b=0.71^{+0.06}_{-0.05}$ (at $1.5 h_{50}^{-1}$ Mpc;
68\% confidence). Hence, the dark matter is more strongly clustered
than the galaxies. The variation of $b$ with scale is significant, but
a better determination of $\langle M_{\rm ap}^2\rangle$ is needed to
study the scale dependence in more detail.

\vbox{
\vspace{-0.3cm}
\begin{center}
\leavevmode 
\hbox{% 
\epsfxsize=\hsize 
\epsffile[50 150 590 720]{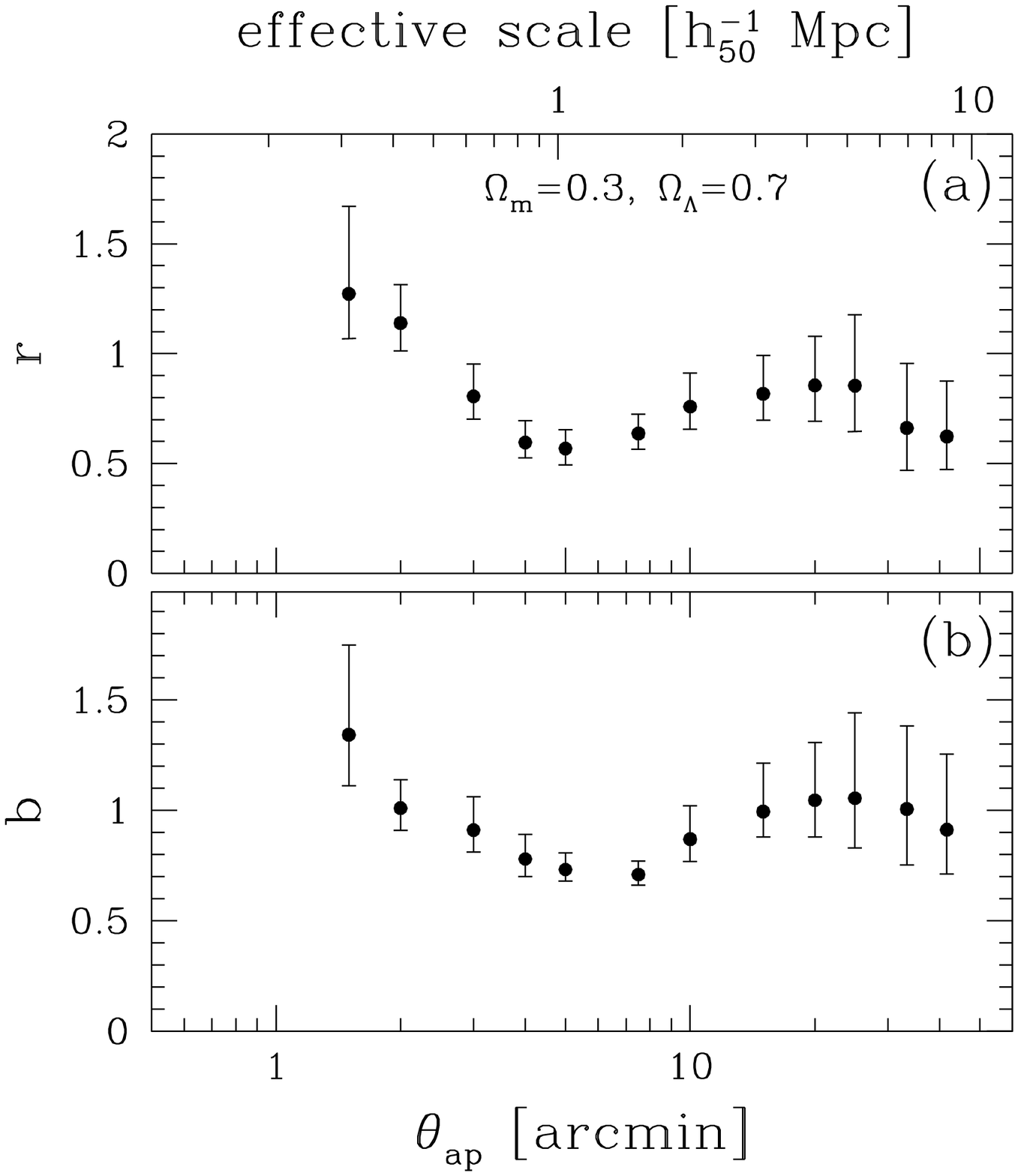}}
\figcaption{\footnotesize (a) The measured value of the galaxy-mass
cross correlation coefficient $r$ as a function of scale for the
$\Lambda$CDM cosmology. (b) The bias parameter $b$ as a function
of scale. The upper axis indicates the effective physical scale probed by the
compensated filter $U(\phi)$ at the median redshift of the lenses
$(z=0.35)$. The errorbars correspond to the 68\% confidence intervals.
Note that the measurements at different scales are slightly
correlated.
\label{bias}}
\end{center}}

Our result is in good qualitative agreement with the findings of
Jenkins et al. (1998) who combined the observed APM two-point
correlation function and the matter correlation function derived from
numerical simulations.  We note that a direct comparison cannot be
made with other measurements (such as those from Jenkins et al. 1998),
because the bias properties depend on galaxy type and redshift. For
instance, observations show that the amplitude of the galaxy
correlation function depends on luminosity (e.g., Benoist et al. 1996;
Norberg et al. 2001).

Our results are derived for a specific subset of galaxies, and are
redshift averaged values (the lenses have a large range in redshift,
with intrinsically brighter galaxies at higher redshifts). This is
clearly demonstrated by Eqn~A7. In the simple case that the average
bias parameters change approximately linearly with redshift, our
results can be interpreted as a measurement of the biasing properties
of $L_*$ galaxies at redshift $\sim 0.35$. However, the redshift
dependence of the bias in magnitude limited samples is usually more
complex.

Hoekstra et al. (2001a) measured the ratio $b/r$ on angular scales out
to 12.5 arcminutes, and found that the results were consistent with a
constant value.  With the additional data (which probe larger scales
and give smaller errorbars) we find evidence for a small trend with
scale. The measurements of the current RCS data extend out to
an angular scale of 1 degree (which corresponds to a physical scale of
$12.5 h_{50}^{-1}$ Mpc), and are presented in Figure~\ref{b_over_r}a.
Figure~\ref{b_over_r} shows that even on a 1 degree scale, the
systematics are much smaller than the signal. Hence, the value of
$b/r$ can be determined accurately. The new data yield an average value of
$b/r=1.090\pm0.035$. Limiting the measurements to the angular scales
studied in Hoekstra et al. (2001a), we find an average value of
$b/r=1.05\pm0.04$, in excellent agreement with the previous estimate of
$b/r=1.05^{+0.12}_{-0.10}$.

Our measurements should be compared to models of galaxy formation.
The two commonly used approaches are hydrodynamic simulations (e.g.,
Blanton et al. 2000, Yoshikawa et al. 2001) or semi-analytical models
(e.g., Kauffmann et al. 1999a,b; Somerville et al. 2001, Guzik \&
Seljak 2001). These studies suggest values of $r$ close to
unity. Unfortunately, the mass resolution in these simulations, such
as the GIF results (e.g., Kauffmann et al. 1999a, b), is too poor:
they only resolve massive, luminous galaxies. Our sample of lenses
contains many lower luminosity systems, which seriously hampers the
comparison of our results with predictions.

As the value of $r$ is intimately linked to the details of galaxy
formation, a careful comparison of the models with the weak lensing
measurements provides unique constraints. In addition, planned large
weak lensing surveys, such as the CFHT Legacy Survey\footnote{\tt
http://www.cfht.hawaii.edu/Science/CFHLS/}, will significantly improve
the accuracy of the measurements, allowing larger scales to be probed,
and enabling us to select galaxies based on the colours or
luminosities (using photometric redshifts). Therefore the prospects of
constraining biasing parameters from weak lensing are excellent.

\vbox{
\vspace{-0.3cm}
\begin{center}
\leavevmode 
\hbox{% 
\epsfxsize=8cm
\epsffile[50 150 590 720]{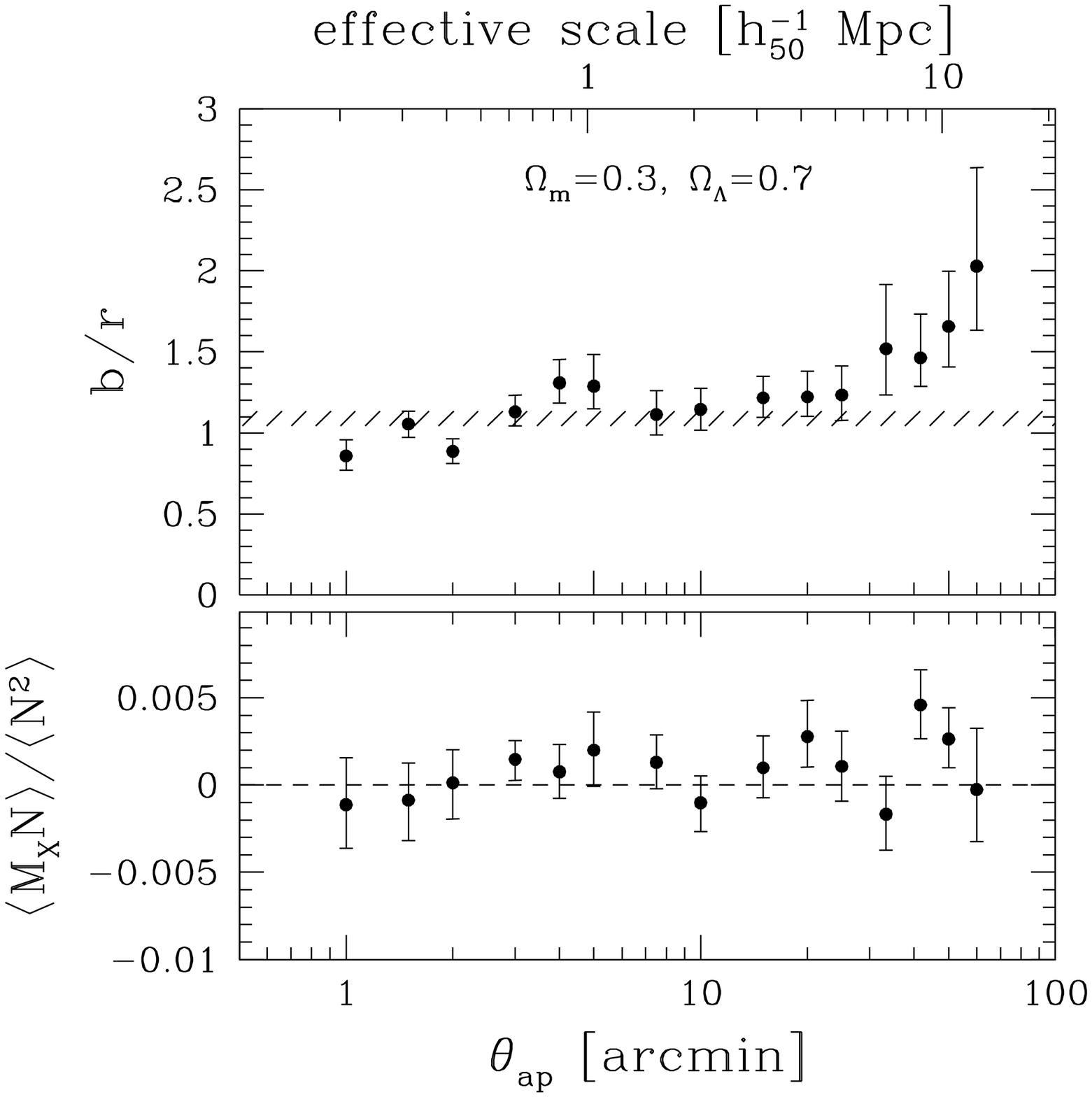}}
\figcaption{\footnotesize (a) The value of $b/r$ as a function of
angular scale, under the assumption $\Omega_m=0.3$ and
$\Omega_\Lambda=0.7$. These results are based on RCS data only, which
allows this ratio to measured out to 1 degree (which corresponds to a
physical scale of $12.5 h_{50}^{-1}$ Mpc). Note that the points are
slightly correlated. The value of $b/r$ is almost constant over the
range probed here. For this cosmology we find an average value of
$b/r=1.090\pm 0.035$, in excellent agreement with the result from
Hoekstra et al. (2001a). (b) Measurement of 
$\langle M_{\rm X}{\cal N}\rangle/\langle{\cal N}^2\rangle$ 
(the signal when the phase of the shear is increased by $\pi/2$),
which should vanish if the results in (a) are caused by lensing.
\label{b_over_r}}
\end{center}}

\acknowledgments We thank the anonymous referee for his comments,
which helped to improve the paper significantly. The RCS is supported
by a NSERC operating grant to HY. We thank the VIRMOS and Terapix
teams who obtained and processed the VIRMOS-DESCART data. This
VIRMOS-DESCART survey was supported by the TMR Network ``Gravitational
Lensing: New Constraints on Cosmology and the Distribution of Dark
Matter'' of the EC under contract No. ERBFMRX-CT97-0172.  YM thanks
CITA for hospitality. It is a pleasure to thank Simon White, Peter
Schneider, and Francis Bernardeau for useful discussions.

\appendix
\section{Scale and redshift dependent bias parameters}

As has been demonstrated in this paper, the bias relation is more
complicated than the simple case of linear deterministic biasing, and
the inferred values of $b$ and $r$ vary with scale. In addition, these
parameters are expected to vary with redshift.

In \S3 we treated both $b$ and $r$ as constants, which is not
warranted by the data. In this appendix, however, we show that as long
as $b$ and $r$ vary slowly with scale, we can still infer the bias
parameters as a function of scale directly. This procedure works,
because the aperture mass statistic is effectively a pass-band filter.

\noindent If we assume that $\delta_g(k)=b(k)\delta(k)$, we have to
replace Eqn.~5 with

\begin{equation}
\Delta n_{\rm g}(\theta)= \int dw~p_f(w)b(f_K(w)\theta;w)\delta(f_K(w)\theta;w),
\end {equation}

\noindent which changes the galaxy auto-correlation function to

\begin{equation}
\langle{\cal N}^2(\theta_{\rm ap})\rangle=2\pi \int dw 
\frac{p_f^2(w)}{f_K^2(w)} \tilde P_{\rm filter}(w;\theta_{\rm ap}),
\end{equation}

\noindent where

\begin{equation}
\tilde P_{\rm filter}(w;\theta_{\rm ap})\hspace{-0.3mm}=\hspace{-1.2mm}\int\hspace{-1.0mm}
dl~l~b^2\left(\frac{l}{f_K(w)};w\right) 
P_{\rm 3d}\hspace{-0.3mm}\left(\frac{l}{f_K(w)};w\right)
J^2(l \theta_{\rm ap}).
\end{equation}

\noindent The filter $J^2(\eta)$ is strongly peaked, and this
motivates the approximation of $J^2$ by a Dirac delta function
(Bartelmann \& Schneider 1999)

\begin{equation}
J^2(\eta)\approx \frac{512}{1155\pi^3}\delta(\eta-693\pi/512)
\approx 1.43\times10^{-2}\delta(\eta-4.25).
\end{equation}

\noindent Consequently $\tilde P_{\rm filter}$ can be approximated as 

\begin{equation}
\tilde P_{\rm filter}(w;\theta_{\rm ap})\approx 
b^2\left[\frac{4.25}{\theta_{\rm ap}f_K(w)};w\right] P_{\rm filter}(w;\theta_{\rm ap}),
\end{equation}

\noindent where $P_{\rm filter}$ is given by Eqn.~9. For the ratio of
the galaxy and matter auto-correlation functions we can write

\begin{equation}
\frac{\langle M_{\rm ap}^2\rangle}{\langle{\cal N}^2\rangle}\approx
\frac{9}{4}\left(\frac{H_0}{c}\right)^4\Omega_m^2
\frac{\int dw h_2(w;\theta_{\rm ap})}{\int dw h_1(w;\theta_{\rm ap})
b^2(4.25/(\theta_{\rm ap}f_K(w));w)}.
\end{equation}

\noindent or

$$
\frac{\int dw h_1(w;\theta_{\rm ap})b^2(4.25/(\theta_{\rm ap}f_K(w));w)}
{\int dw h_1(w;\theta_{\rm ap})}=
\frac{9}{4}\left(\frac{H_0}{c}\right)^4\Omega_m^2
\frac{\int dw h_2(w;\theta_{\rm ap})}{\int dw h_1(w;\theta_{\rm ap})}
\frac{\langle{\cal N}^2\rangle}{\langle M_{\rm ap}^2\rangle}\nonumber
$$
\begin{equation}
 = f_1\times \Omega_m^2 \times
\frac{\langle{\cal N}^2\rangle}{\langle M_{\rm ap}^2\rangle}(\theta_{\rm ap}).
\end{equation}

\noindent The functions $h_1$, and $h_2$ are given by Eqns.~8 and~12.
The actual calculations show that $f_1$ does not depend on the power
spectrum (e.g., van Waerbeke 1998).  The left hand side of Eqn.~A7
shows that we measure the bias parameter weighted by the function
$h_1$. 

A similar result can be obtained for the galaxy-mass cross-correlation
coefficient $r$

$$
\frac{\int dw h_3(w;\theta_{\rm ap})r(4.25/(\theta_{\rm ap}f_K(w));w)}
{\int dw h_3(w;\theta_{\rm ap})}=
\frac{\sqrt{\int dw h_1(w;\theta_{\rm ap})}\sqrt{\int dw h_2(w;\theta_{\rm ap})}}
{\int dw h_3(w;\theta_{\rm ap})}\times\frac{\langle M_{\rm ap}{\cal N}\rangle}
{\sqrt{\langle{\cal N}^2\rangle \langle{M_{\rm ap}^2\rangle}}}.
$$

\begin{equation}
 = f_2\times \frac{\langle M_{\rm ap}(\theta_{\rm ap})\cal{N}(\theta_{\rm ap})\rangle}
{\sqrt{\langle {\cal N}^2(\theta_{\rm ap})\rangle
\langle M_{\rm ap}^2(\theta_{\rm ap})\rangle}}.
\end{equation}

The redshift distribution of the lenses $p_f(w)$ is narrow (see
Fig.~\ref{z_probe}).  Hence, we measure the bias parameters over a
small range in redshift. Because $f_K(w)$ varies somewhat over this
interval, we average values of $b$ and $r$ on different physical
scales. However, if $b$ and $r$ vary slowly with scale and redshift,
we can ignore the variation of these parameters over this range in
scale. Hence, Eqns.~12 and~14 can be used to obtain a direct
measurements of the bias parameters as a function of scale.

\end{document}